\documentclass[%
twocolumn,
superscriptaddress,
showpacs,preprintnumbers,
 amsmath,amssymb,
 aps,
prl,
]{revtex4}

\usepackage{graphicx}% Include figure files
\usepackage{dcolumn}% Align table columns on decimal point
\usepackage{bm}% bold math
\usepackage{color} % change color of text
\usepackage{CJK}

\def\la{{\langle}}
\def\ra{{\rangle}}

\newcommand{\beq}{\begin{equation}}
\newcommand{\eeq}{\end{equation}}
\newcommand{\beqa}{\begin{eqnarray}}
\newcommand{\eeqa}{\end{eqnarray}}

\begin{document}
%\begin{CJK*}{GB}{}
\title{Fast and robust spin manipulation in a quantum dot by electric fields}

\author{Yue Ban}
\affiliation{Departamento de Qu\'{\i}mica-F\'{\i}sica, UPV/EHU, Apdo 644, 48080 Bilbao, Spain}

\author{Xi Chen}
\affiliation{Departamento de Qu\'{\i}mica-F\'{\i}sica, UPV/EHU, Apdo 644, 48080 Bilbao, Spain}
\affiliation{Department of Physics, Shanghai University, 200444 Shanghai, People's Republic of China }

\author{E. Ya Sherman}
\affiliation{Departamento de Qu\'{\i}mica-F\'{\i}sica, UPV/EHU, Apdo 644, 48080 Bilbao, Spain}

\affiliation{IKERBASQUE, Basque Foundation for Science, 48011 Bilbao, Spain}

\author{J. G. Muga}
\affiliation{Departamento de Qu\'{\i}mica-F\'{\i}sica, UPV/EHU, Apdo 644, 48080 Bilbao, Spain}

\affiliation{Department of Physics, Shanghai University, 200444 Shanghai, People's Republic of China }

\begin{abstract}
We apply an invariant-based inverse engineering method to control
by time-dependent electric fields  electron spin  dynamics
in a quantum dot with spin-orbit coupling in a weak magnetic field.
The designed electric fields provide a shortcut to adiabatic processes
that flips the spin rapidly, thus avoiding decoherence effects. This approach, being robust with respect
to the device-dependent noise, can open new possibilities for the spin-based quantum information processing.
\end{abstract}
%73.63.Kv Quantum dots
%72.25.Dc Spin polarized transport in semiconductors
%72.25.Pn Current-driven spin pumping
%}
\maketitle
%\end{CJK*}
%
%
%*********************************************************************
%
%\section{Introduction}
%
%
Coherent spin manipulation in quantum
dots (QDs) \cite{spin resonance,control-geometric phase1,control-geometric phase2,optical-control1,optical-control2,optical-control3,Nowack,Petta,Giroday}
is the key element in the state-of-the-art spintronics and solid-state quantum information \cite{Hanson,Dyakonov}.
Accurate spin manipulation can be achieved by several techniques.
One of them is the conventional electron spin resonance
induced by a magnetic field oscillating at the Zeeman transition frequency  \cite{spin resonance}.
%To achieve fast and precise spin control, there are several approaches including the conventional electron spin resonance
%(ESR) that induces transitions of a
%localized electron spin with a Zeeman-resonant oscillating magnetic field \cite{spin resonance}.
A more robust technique is the spin manipulation with geometric Berry phases during adiabatic motion
\cite{control-geometric phase2, control-geometric phase1}. Nowadays,
there is also a growing interest in the electric control of
spin using spin-orbit (SO) coupling \cite{Rashba}. It has been applied to high-fidelity
spin manipulation on the $100$ ns time scale \cite{Nowack,Petta,Giroday}.
This highly efficient all-electrical method has several advantages.
For example, it is easy to generate time-dependent electric fields on the nanoscale by
adding local electrodes and produce spin manipulation by making them
Zeeman-resonant \cite{Nowack}. As a result, Rabi spin oscillations appear at a frequency
much smaller than the Zeeman frequency making the flip relatively slow and prone to decoherence.
We shall propose here another all-electrical technique to flip spin
with high fidelity via ``shortcuts to adiabaticity'',
in a time that can be much shorter than any decoherence time.

%More easily generated than magnetic field experimentally, electric field in the manipulations of electron
%spin \cite{Nowack,Petta,Giroday} possesses other advantages in terms of coupling electron spin in the quantum structures simply through
%On the other hand, due to the major difficulty of the strong hyperfine interaction which couples the electron spin to a bath of nuclear spin, studies of hole
%spins with suppressions of nuclear feedback by means of optical controls have been motivated \cite{optical-control1,optical-control2,optical-control3}.

Recently, several shortcuts to adiabaticity have been put forward to
speed up the adiabatic passage of quantum systems,
and achieve a robust and fast adiabatic-like control
\cite{Rice,Berry09,Chen10b,Oliver,bec,Chen,Nice,Nice2,3d,ChenPRA,Masuda,Adol,Onofrio,transport}.
The transitionless or counter-diabatic control algorithms proposed by Demirplak, Rice \cite{Rice}, and Berry \cite{Berry09},
are designed to add supplementary time-dependent interactions that cancel the diabatic couplings of a reference process.
The system then follows exactly the adiabatic trajectory of the original unperturbed process, in principle in an
arbitrarily short time. Transitionless quantum drivings
have been implemented in two-level systems: spins in a magnetic field \cite{Berry09}, atoms \cite{Chen10b},
and Bose-Einstein condensates in optical lattices \cite{Oliver}. A different shortcut  is provided by inverse
engineering the transient Hamiltonian \cite{bec,Chen} using Lewis-Riesenfeld invariants \cite{LR}.
This method has been used for time-dependent traps
\cite{bec,Chen,Nice,Nice2,3d}, atomic transport \cite{transport},
and other applications \cite{Adol,Onofrio}. Although these two methods are potentially equivalent \cite{ChenPRA},
their implementations and results can be quite different. Here we choose the invariant-based inverse
engineering approach, since it is better suited than the transitionless driving to be produced by the desired
all-electrical means.
%that controls electron spin in a quantum dot
%with spin-orbit coupling. Fields are designed to implement a high-fidelity spin flip in a short time much faster than the possible decoherence effects.  %\cite{dephasing-time1,dephasing-time2}.
%so that the speeded-up process is not influenced by the dephasing.

\begin{figure}[]
\scalebox{0.50}[0.50]{
\includegraphics{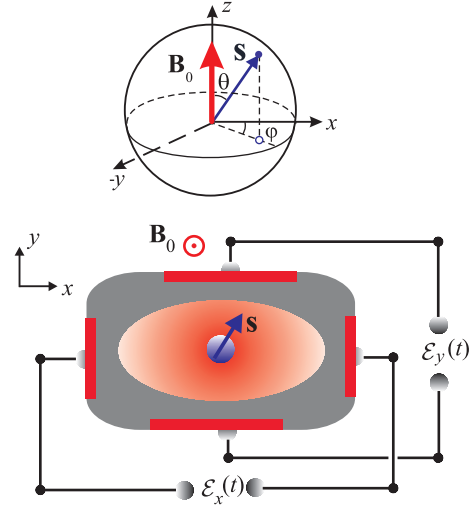}}
\caption{(Color online) Schematic diagram of spin dynamics of electron in a QD in the presence of electric fields
$\mathcal{E}_{i}(t)$, and perpendicular magnetic field $\mathbf{B}_0\parallel{z}-$axis.}
\label{model}
\end{figure}

\textit{Model.-}
We consider the electric control of electron spin in a QD formed in the $x$-$y$ plane of a two-dimensional electron gas
confined in the $z$-direction by the coordinate-dependent material composition, under
a weak magnetic field $\mathbf{B}_0 \parallel \mathbf{z}$, as shown in Fig. \ref{model}.
%%%%%%%%%%
%%%%%%%%%%
Here the total Hamiltonian $H$ of the electron interacting with the external electric field
${\bf \mathcal{E}} (t) = -\partial \textbf{A}/c\partial t$ is
$
H=H_0+ H_{\rm so} + H_{\rm int},
$
with \cite{Rashba}
\beqa
\label{H_0}
H_0 &=& \frac{p^2_x+p^2_y}{2m} + U(x,y) + \frac{\Delta_z}{2} \sigma_z ,
\\
H_{\rm so} &=& (-\alpha \sigma_y  +  \beta \sigma_z) p_x + \alpha\sigma_x p_y,
\\
H_{\rm int} &=& - \frac{e}{c} \textbf{A}(t) \cdot \textbf{v},
\eeqa
where $m$ is the electron effective mass and $\sigma_i$ ($i=x,y,z$) are the Pauli matrices.
$H_0$ represents the kinetic energy, the
potential $U(x,y)$, and the Zeeman splitting $\Delta_z = g \mu_B B_0$, where
$\mu_B$ is the Bohr magneton, and $g$ is the Land\'{e} factor. The eigenfunctions of $H_0$ are $\psi_{j}(x,y)\left|\sigma\right>$,
where $|\sigma\ra=|\pm1\ra$ is the eigenstate of $\sigma_{z}$, and the spectrum is given by $E_{j}\pm\Delta_{z}/2$, where
$E_{j}$ are the orbital eigenenergies in the confinement potential.

The SO coupling is the sum of structure-related Rashba ($\alpha$)  and bulk-originated
Dresselhaus ($\beta$) terms for $[1 1 0]$ growth axis \cite{Golovach}.
The vector potential $\textbf{A}(t)$ is in the $(x,y)-$plane and corresponding
spin-dependent velocity operators are
\beqa
v_x  &=& \frac{i}{\hbar} \left[H_0+H_{\rm so}, x\right] = p_x/m + \beta \sigma_z  - \alpha \sigma_y,
\\
v_y  &=& \frac{i}{\hbar} \left[H_0+H_{\rm so}, y\right] = p_y/m + \alpha \sigma_x.
\eeqa
%

%To describe the realistic electron spin dynamics in such quantum dot, taking into account the influence of orbital motion and excited energy states, we
%firstly ignore the higher excited states and retain the four lower-energy eigenstates of Hamiltonian $H_0$,
%$| \psi_1 \ra = \Psi_1 (x) |1\ra$,  $| \psi_2 \ra = \Psi_1 (x) |0\ra$,
%$| \psi_3 \ra = \Psi_2 (x) |1 \ra$, and $| \psi_4 \ra = \Psi_2 (x) |0 \ra$,
%where $\Psi_{1,2}$ are the wave functions of orbital motions with the kinetic energies $E_1$ and $E_2$, and
%$|1 \ra ={1\choose 0}$ and $|0 \ra={0\choose 1}$ are spinors.

We focus on the doublet $\Psi_{1}=\psi_1\left|1\right>$,$\Psi_{2}=\psi_1\left|-1\right>$,
include the higher orbitals by  L\"{o}wdin partition \cite{lowdin-partition,Winkler}, and
reduce the full Hamiltonian into an effective $2 \times 2$ one \cite{suppl material},
\beqa
\label{H'}
H^{\textrm{eff}} = \frac{g \mu_B}{2}
\left(
\begin{array}{cc}
Z & X + i Y
\\
X - i Y & -Z
\end{array}
\right),
\eeqa
where $X = B_2(1+\xi_y) $, $Y = (\alpha/\beta)(1+\xi_x) B_1 $, $Z = B_0 + (1+\xi_x) B_1 $,
with the effects of higher states characterized by $\xi_x$ and $\xi_y$, and
the components of the designed electric field are renormalized by the factors of $1/(1+\xi_{i})$.
The effective magnetic fields are expressed with the SO coupling
parameters as $B_1= - 2 e \beta A_x/c g \mu_B$, and $B_2=- 2 e \alpha A_y/c g \mu_B$.
The resulting electric fields are:
\beqa
\label{electric field}
\mathcal{E}_x (t)= \frac{g \mu_B  }{2 e \beta } \frac{\partial B_1}{\partial t},~~
\mathcal{E}_y (t)= \frac{g \mu_B  }{2 e \alpha } \frac{\partial B_2}{\partial t}.
\eeqa
%
%which are determined by the external fields.
In practice, some slowly varying electric fields can be applied to drive the state from $\Psi_1$ to $\Psi_2$
adiabatically along an instantaneous eigenstate of
the Hamiltonian in Eq. (\ref{H'}). To accelerate the driving using the transitionless algorithm,
counter-diabatic fields should be provided \cite{ChenPRA}.
However, the common dependence of $Y$ and $Z$ on $B_1$ precludes the
implementation of the fast driving terms only by electric fields.
In contrast, invariant-based inverse engineering naturally leads to
an all-electrical driving.

\textit{Dynamical invariant and spin-flip example.-}
We shall design the time dependence of the external electric fields to guarantee
the state transfer in some fixed time $t_f$ by using the dynamical $2\times 2$ invariant $I(t)$
satisfying the condition $dI(t)/dt \equiv \partial I(t)/\partial t  - [H^{\textrm{eff}}(t), I(t)]/ i \hbar=0$.
Parametrizing the Bloch sphere (Fig. \ref{model}),
by the angles $\theta$ and $\varphi$, we construct
yet unknown orthogonal eigenstates $|\chi_{\pm} (t) \ra$ of $I(t)$ as
%%%%%%%%%%%%%%%%%%%%%%%%%%%%%%
\beqa
|\chi_{+}(t) \ra
=
\left(\begin{array}{c}
\cos\displaystyle{\frac{\theta}{2}}  e^{i \varphi}
\\
\sin\displaystyle{\frac{\theta}{2}}
\end{array}
\right),
|\chi_{-}(t)  \ra
=
\left(\begin{array}{c}
 \sin \displaystyle{\frac{\theta}{2}}
\\
- \cos \displaystyle{\frac{\theta}{2}} e^{-i \varphi}
\end{array}
\right).~
\eeqa
They satisfy $I(t) |\chi_\pm (t)\ra = \lambda_\pm |\chi_\pm (t)\ra$.
%%%%%%%%%%%%%%%%%%%%%%%%%%%%%%
Introducing $\lambda_{\pm}=\pm g \mu_B B_{c}/2$,
we construct the invariant
%$I(t)=\sum_n \lambda_n |\chi_n (t) \ra \la \chi_n (t) | $
as \cite{ChenPRA}
\beqa
\label{invariant}
I (t) = \frac{g \mu_B }{2} B_{c}
\left(
\begin{array}{cc}
\cos{\theta} & \sin{\theta} e^{i \varphi}
\\
\sin{\theta} e^{-i \varphi} & -\cos{\theta}
\end{array}
\right),
\eeqa
where
$B_{c}$ is an arbitrary constant magnetic field to keep $I(t)$ with units of energy. According to the
Lewis-Riesenfeld theory,
the solution of the Schr\"{o}dinger equation, $i \hbar  \partial_t \Psi=H^{\textrm{eff}}(t) \Psi$, is a superposition
of orthonormal ``dynamical modes", $\Psi (t) = \sum_n C_n e^{i \alpha_{n}(t)} |\chi_{n} (t) \ra$ \cite{LR}, where
$C_n$ are time-independent amplitudes and
$\alpha_{n}(t)$ is Lewis-Riesenfeld phase,
\beq
\alpha_n (t) =\frac{1}{\hbar} \int^t_0 \langle \chi_n(t') | i\hbar \frac{\partial }{\partial t'} - H^{\textrm{eff}}(t')| \chi_n(t') \rangle dt'.
\eeq
From the invariant condition, $dI(t)/dt=0$,
the angles $\theta$ and $\varphi$ are related to $X$, $Y$ and $Z$ by
auxiliary equations
\beqa
\label{auxiliary equations}
\dot{\theta}  &=&  \eta (X \sin\varphi - Y \cos\varphi),
\\
\label{auxiliary equations2}
\dot{\varphi} &=&  \eta (X \cos\varphi \cot\theta + Y \sin\varphi \cot\theta - Z),
\eeqa
where $\eta =g \mu_B/\hbar$. Since $X$ is a function of $B_2$,
while $Y$ and $Z$ are functions of $B_1$, once $\theta$ and $\varphi$ are fixed,
Eqs. (\ref{auxiliary equations}) and
(\ref{auxiliary equations2}) give the effective magnetic fields
\beqa
\label{B_1}
B_1 &=& \frac{ -\beta \dot{\theta} \cot\theta \cos\varphi + \beta (\dot{\varphi} + \eta B_0) \sin\varphi}{ \eta (1+\xi_x)(\alpha \cot\theta -\beta\sin\varphi)},
\\
\label{B_2}
B_2 &=& \frac{\alpha \dot{\theta} \cot\theta \sin\varphi + \alpha (\dot{\varphi} + \eta B_0) \cos\varphi - \beta \dot{\theta}}{\eta (1+\xi_y) (\alpha \cot\theta -\beta\sin\varphi)},
\eeqa
from which the electric fields are calculated using Eq. (\ref{electric field}).
%since $X$, $Y$ and $Z$ in the Hamiltonian (\ref{H'}) are only dependent of $B_1$ and $B_2$,
During the spin-flip process, there exist some time instants $t=t_s$ which satisfy
\beq
\label{denominator}
\alpha \cot \theta(t_s) = \beta \sin \varphi(t_s),
\eeq
and make the denominators of $B_1$ and $B_2$ zero.  To get rid of such singularities we impose the conditions
\beqa
\label{numerator B_1}
\beta \sin \varphi(t_s)\!\left[\dot{\varphi}(t_s) + \eta B_0 -(\beta/\alpha) \dot{\theta} (t_s) \cos\varphi (t_s)\right]\!=\! 0, ~~~
\\
\label{numerator B_2}
\alpha\cos \varphi(t_s)\!\left[\dot{\varphi}(t_s) + \eta B_0 - (\beta/\alpha) \dot{\theta}(t_s)\cos\varphi(t_s)\right]\!=\! 0,~~~
\eeqa
which make the numerators of $B_1$ and $B_2$ zero simultaneously. In the following example, we will show how this works.

%%%%%%%%%%%%%%%%%%%%%%%%%%%%%%%%%%%%%%%%%%%%%%%%%%%%%%%%%%%%%%%%%%%%%%%%%%%%%%%%%%%%%%%%%%%%%%%%%%%%%%%%%%%%%%%%%
\begin{figure}[]
\scalebox{0.60}[0.60]{\includegraphics{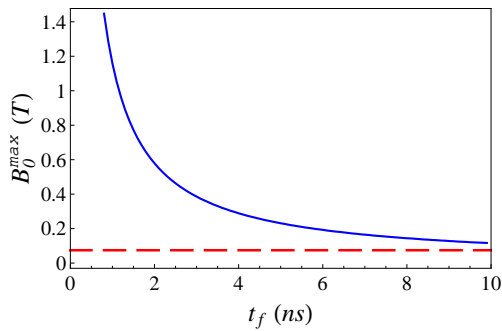}}
\caption{(Color online) Dependence of the maximum of applied magnetic
field $B^{\max}_0$ (solid blue) on the time $t_f$ for a third order polynomial ansatz for $\theta$ and $\varphi$,
with the parameters:
$\hbar \alpha=2 \times 10^{-6}$ meV$\cdot$cm, $\beta = \alpha/2$, $g=-0.44$ for GaAs. $B_0=0.075$ T
(dashed red) corresponds to $\Delta_z= 23$ mK.}
\label{Bmax}
\end{figure}
%%%%%%%%%%%%%%%%%%%%%%%%%%%%%%%%%%%%%%%%%%%%%%%%%%%%%%%%%%%%%%%%%%%%%%%%%%%%%%%%%%%%%%%%%%%%%%%%%%%%%%%%%%%%%%%%%%

%Our aim is to realize the spin flip by using dynamical invariant for a fixed short time.
In general, the eigenstates of the invariant are not the same as the instantaneous eigenstates of the Hamiltonian,
since $I(t)$ and $H^{\rm eff}(t)$ do not commute.
If we impose for $\theta$ at $t=0$ and $t_f$ that
\beqa
\label{boundary1}
\theta (0) = 0, ~ \theta (t_f) = \pi, ~
\dot{\theta}(0) = 0,~ \dot{\theta}(t_f) = 0,
\eeqa
then $[H^{\textrm{eff}}(0), I(0)]=0$ and $[H^{\textrm{eff}}(t_f), I(t_f)]=0$,
which guarantees
common eigenstates at initial and final times.
Moreover the state obeying Eq. (\ref{boundary1}) will flip from $|\Psi_{1}\ra$ at $t=0$ to $|\Psi_{2}\ra$ at $t=t_f$, up to phase factors, along the eigenstate $|\chi_{+}(t) \ra$.
To design the trajectory at intermediate times
we assume the polynomial ansatz $\theta=\sum_{j=0}^3 a_j t^j$, where the $a_j$ can be fixed by solving the system implied by  Eq. (\ref{boundary1}).
This leads to $\theta(t_f/2)=\pi/2$, so $\cot \theta$ covers the whole $(-\infty,\infty)$ range
passing through zero at $t=t_f/2$. This may lead to one or several times satisfying Eq. (\ref{denominator}), as we will see below in more detail.

%However, we still have freedom to choose $\varphi$, which is irrelevant to achieve commutativity,
%because the initial and final states are bare states.
To determine $|\chi_{+} (t)\ra$ fully, we also need the trajectory for $\varphi$.
As the initial and final states
are the poles of the Bloch sphere, the phase $\varphi$ is not well defined there.
We may nevertheless specify how the trajectory approaches them, and impose
limits from the right at $t=0$, and from the left at $t=t_f$, for example,
\beqa
\label{boundary2}
\varphi (0^+) = \pi/2, ~~~~ \varphi (t^{-}_f) = \pi/2.
\eeqa
These conditions are not sufficient though, since we still have to deal with the singularities and their cancellation.
%From Eq. (\ref{denominator}), we know there could be several singularities at $t=t_s$,
%when $\sin \varphi(t_s) = (\alpha/\beta) \cot\theta(t_s)$ is satisfied.
As $\cot \theta(t_f/2) =0$, we may satisfy Eq. (\ref{denominator})
and impose zeros of the denominators for $B_{1,2}$ at $t_s=t_f/2$, if
$\sin \varphi(t_f/2) =0$. Imposing the two conditions
%We would like to add the following two conditions at $t=t_f/2$,
%
\beqa
\label{boundary3}
\varphi(t_f/2) = 0, ~~~~~~~~~~
\\
\label{boundary4}
\dot{\varphi}(t_f/2) = (\beta/\alpha) \dot{\theta}(t_f/2)- \eta B_0,
\eeqa
to satisfy {Eqs. (\ref{numerator B_1}) and (\ref{numerator B_2})}
at $t_s=t_f/2$, we
cancel the singularity there.
%Eq. (\ref{boundary3}) and $\cot \theta(t_f/2)=0$ result in zero value of the denominators of $B_1$ and $B_2$ at $t_f/2$. Meanwhile, denominator of $B_1$ is also zero at this time instant so %that the singularity can be avoided in $B_1$. However, for $B_2$, additional condition Eq. (\ref{boundary4}) is needed.
With the conditions in Eqs. (\ref{boundary2})-(\ref{boundary4}), we solve the third order polynomial
ansatz $\varphi=\sum_{j=0}^3 b_j t^j$ to determine $\varphi(t)$.

Explicit calculations demonstrate that for the third-order polynomial ansatz
and boundary conditions imposed here,
there is only one (removable) singularity at $t_s=t_f/2$  when the field $B_0$ is smaller than certain upper limit $B_0^{\max}$, shown in
Fig. \ref{Bmax} as a function of $t_f$.
%The singularity can solved by $\alpha \cot \theta = \beta \sin \varphi$, once $\alpha$ and $\varphi$ are fixed.
For $B_0>B_0^{\max}$, more solutions of (\ref{denominator})
appear [$B_0$ and $\sin\varphi(t)$ are coupled by
Eq. (\ref{boundary4})],
which cannot be canceled with the third order polynomial. To satisfy
Eqs. (\ref{numerator B_1}) and (\ref{numerator B_2}) at more than one zero of the denominators of $B_{1,2}$, one may set
higher order polynomials for $\varphi$ and further conditions. This increases the bound $B_0^{\max}$, but also complicates the driving fields.
%
%%%%%%%%%%%%%%%%%%%%%%%%%%%%%%%%%%%%%%%%%%%%%%%%%%%%%%%%%%%%%%%%%%%%%%%%%%%%%%%%%%%%%%%%%%%%%%%%%%%%%%%%%%%%%%%%%%%%%%%%%%%%%%%%%%%%%%%%%%%%%%%%%%
\begin{figure}[]
\scalebox{0.60}[0.60]{\includegraphics{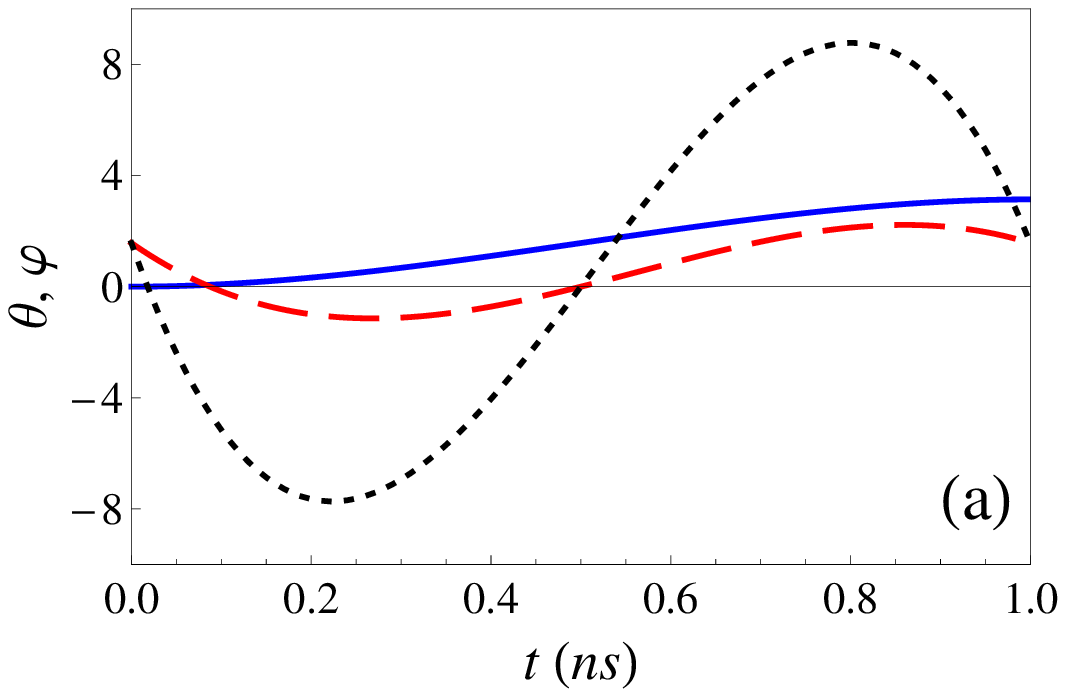}}
\scalebox{0.60}[0.60]{\includegraphics{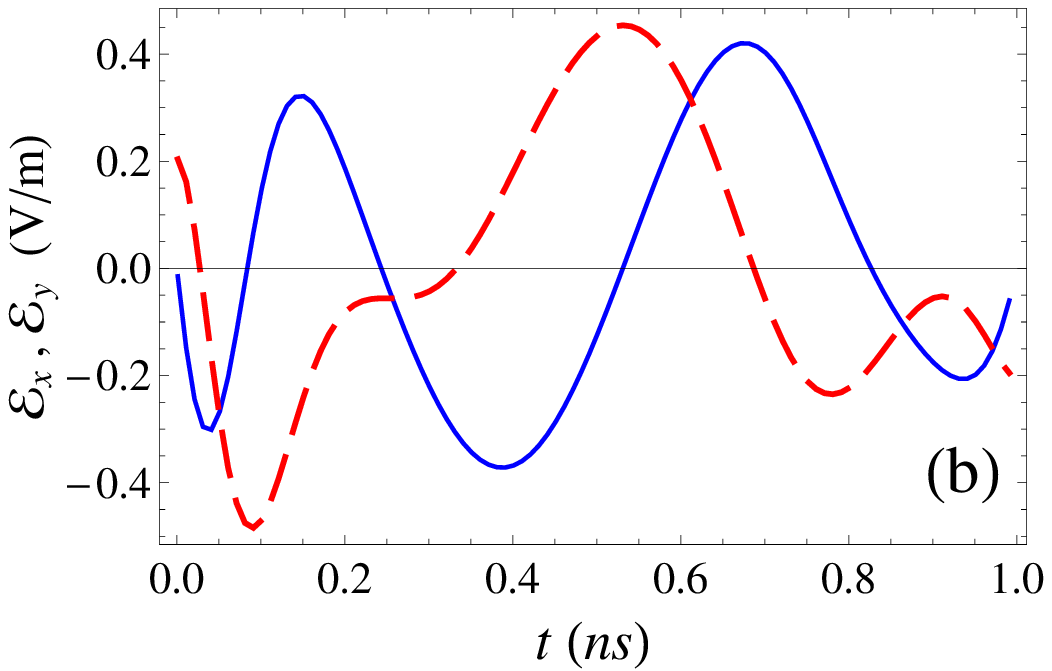}}
\scalebox{0.60}[0.60]{\includegraphics{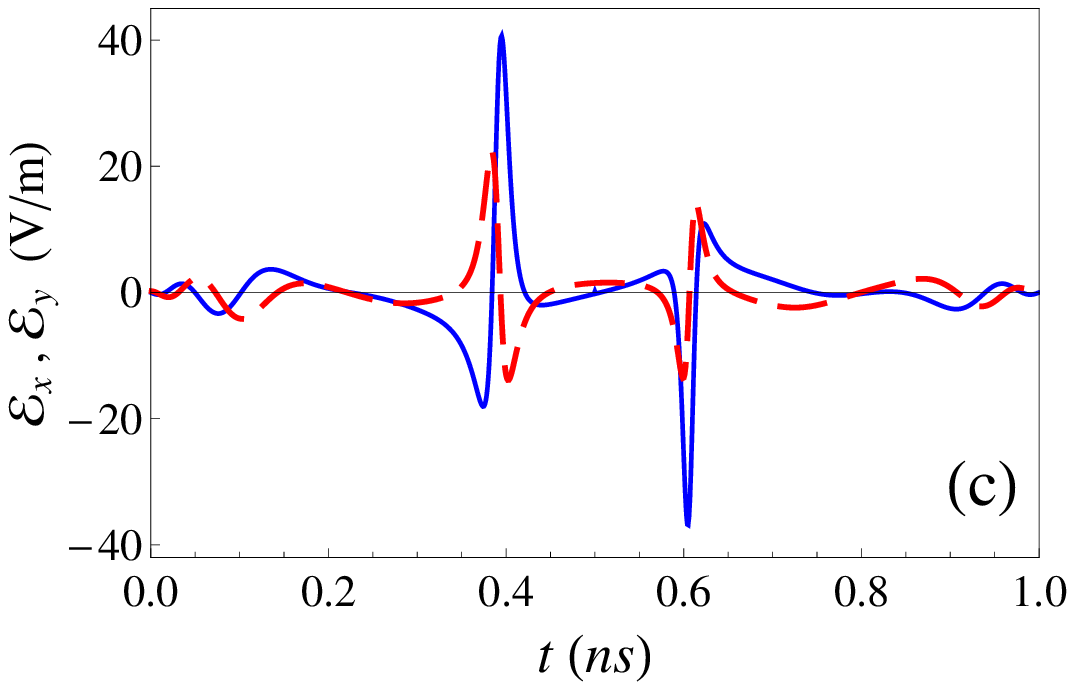}}
\caption{(Color online) For $t_f=1$ ns: (a) Polynomial ansatzes of auxiliary angles $\theta=\sum_{j=0}^3 a_j t^j$ (solid blue),
$\varphi=\sum_{j=0}^3 b_j t^j$ with $B_0=0.15$ T (dashed red) and $B_0=1.05$ T (dotted black).
The designed electric fields ${\mathcal E}_x$ (solid blue) and ${\mathcal E}_y$ (dashed red) by which spin flip
can be realized for $B_0=0.15$ T (b) and $B_0=1.05$ T (c).
Other parameters are the same as those in Fig. \ref{Bmax}.
For simplicity, we put here $\xi_{x}=\xi_{y}=0$ to skip the trivial dependence on these factors.}
\label{example}
\end{figure}
%%%%%%%%%%%%%%%%%%%%%%%%%%%%%%%%%%%%%%%%%%%%%%%%%%%%%%%%%%%%%%%%%%%%%%%%%%%%%%%%%%%%%%%%%%%%%%%%%%%%%%%%%%%%%%%%%%%%%%%%%%%%%%%%%%%%%%%%%%%%%%%%%%
In the present examples, we just apply the third order polynomial ansatz with the boundary conditions
in Eqs. (\ref{boundary1})-(\ref{boundary4}), so that the applied magnetic field $B_0$ should not
go beyond the upper limit in Fig. \ref{Bmax}. As the upper limit field grows for smaller times $t_f$,
this is not a problem in practice. Figure \ref{example} shows  examples of spin flip for different values of $B_0$.

With the functions $\theta$ and $\varphi$ fixed [see Fig. \ref{example} (a)], the designed electric
fields, $\mathcal{E}_x(t)$ and $\mathcal{E}_y(t)$, corresponding to $B_0=0.15$ T and $B_0=1.05$ T
(close to the upper limit), are depicted in Fig. \ref{example} (b) and (c). The populations
(not shown) of the two spin states, given by  $P_{1}= \cos^2{(\theta/2)}$ and $P_{-1}= \sin^2{(\theta/2)}$,
cross each other smoothly as $\theta$ goes from $0$ to $\pi$. The choice of $B_0$ determines the trajectory on the Bloch sphere for a given $t_f$.
When $B_0$ approaches the upper limit, the electric fields exhibit sharp peaks, see Fig. \ref{example} (c).
The smooth time-dependence in Fig. \ref{example} (b) is well suited for the applications,
while the complicated dependence in Fig. \ref{example} (c) should be avoided.
Undesirable excitation of the orbital modes does not occur here
since the spin flip $t_{f}\sim 1$ ns, while the energy split
of the orbital states in typical QDs exceeds $0.1$ meV. Therefore, regarding the orbital
motion, our perturbation is strongly adiabatic, and no orbital excitation occurs.

%%%%%%%%%%%%%%%%%%%%%%%%%%%%%%%%%%%%%%%%%%%%%%%%%%%%%%%%%%%%%%%%%%%%%%%%%%%%%%%%%%%%%%%%%%%%%%%%%%%%%%%%
\begin{figure}[]
\scalebox{0.60}[0.60]{\includegraphics{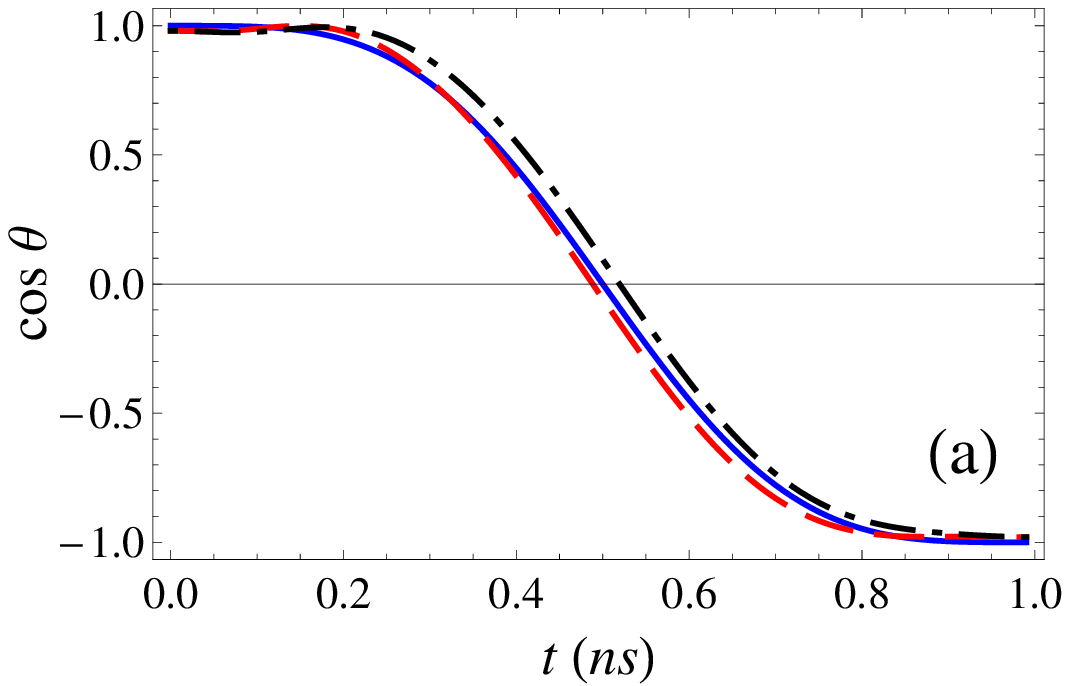}}
\scalebox{0.60}[0.60]{\includegraphics{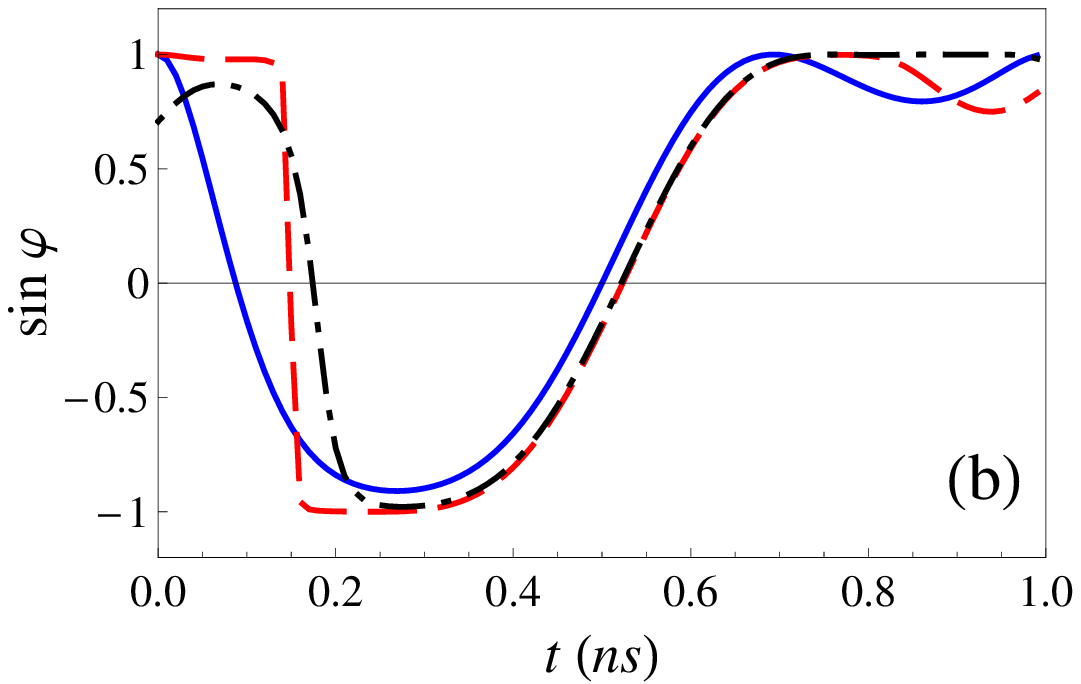}}
\caption{(Color online) Time evolution of $\cos \theta$ (a) and $\sin \varphi$ (b) for the same Hamiltonian with the designed electric fields and $B_0=0.15$ T, where $\epsilon=0$, $\varphi_0=\pi/2$ (solid blue); $\epsilon=0.01$, $\varphi_0=\pi/2$ (dashed red);
$\epsilon=0.01$, $\varphi_0=\pi/4$ (dot-dashed black). Other parameters are the same as in Fig. \ref{example}.}
\label{angles}
\end{figure}
%%%%%%%%%%%%%%%%%%%%%%%%%%%%%%%%%%%%%%%%%%%%%%%%%%%%%%%%%%%%%%%%%%%%%%%%%%%%%%%%%%%%%%%%%%%%%%%%%%%%%%%%

%The electric fields designed so far depend on the polynomial ansatz and boundary conditions (\ref{boundary1})-(\ref{boundary3}).
%In this example, the boundary conditions guarantee that the initial and final states are the bare states with the phase factors, that is, $ \Psi(0)=e^{i \pi/2} |1 \ra $
%and $\Psi(t_f)= e^{i \pi/2 } |0 \ra$.
To check the stability with respect to initialization errors, we assume
now the initial state as $(\sqrt{1-\epsilon} e^{i \varphi_0}, \sqrt{\epsilon})^T$
with an arbitrary phase $\varphi_0$ and find $\theta$ and $\varphi$ from
Eqs. (\ref{auxiliary equations}) and (\ref{auxiliary equations2}) for
the same designed electric fields (Fig. \ref{angles}).
%These results present the time evolution of different initial states, driven by the same Hamiltonian.
%Figure \ref{angles} shows that
The final value
$\theta(t_f)$ depends on the error $\epsilon$, but insensitive to the
initial phase $\varphi_0$,
while $\varphi(t_f)$ is sensitive to both initial conditions. Since our goal
is to realize the spin flip, the final $\varphi$ is irrelevant,
and the experimental effort should focus on achieving a small error $\epsilon$.
%Thus, the electric fields designed here are applicable for the initial states with various phase $\varphi_0$,
%especially when $\epsilon$ is small enough. We emphasize that $\theta(0) =0$ and $\theta(t_f) =\pi$ are necessary conditions
%for the spin flip, and more boundary conditions for $\varphi$ may lead to the
%higher energy consumption per transition.

%If we consider there existing non-purity of bare states at initial time, expressed by $(\sqrt{1-\epsilon} e^{i \varphi}, \epsilon)^T$, we can check the stability of our method by using the designed electric field obtained from the initial state $|1 \ra ={1\choose 0}$, where $\epsilon$ is small enough and $\varphi$ can be any value.

%%%%%%%%%%%%%%%%%%%%%%%%%%%%%%%%%%%%%%%%%%%%%%%%%%%%%%%%%%%%%%%%%%%%%%%%%%%%%%%%%%%%%%%%%%%%%%%%%%%%%%%%%%%%%%%%%%
%\begin{figure}[]
%\scalebox{0.4}[0.4]{\includegraphics{fig5.eps}}
%\caption{(Color online) Time evolution of the Bloch vector when the
%constant magnetic field $B_0=1.05$ T (solid red line) and $B_0=0.15$ T (dashed blue line), where $\gamma=5 \times 10^{7} $ s$^{-1}$  and the other parameters are the same as in Fig. %\ref{example}.}
%\label{bloch}
%\end{figure}
%%%%%%%%%%%%%%%%%%%%%%%%%%%%%%%%%%%%%%%%%%%%%%%%%%%%%%%%%%%%%%%%%%%%%%%%%%%%%%%%%%%%%%%%%%%%%%%%%%%%%%%%

\textit{Decoherence and noise effects.-}
To show feasibility of our approach, we study the effects of noise
and decoherence on the spin-flip fidelity. We begin with a generic approach for coupling
to the incoherent environment, based on the conventional Lindblad formalism as can arise, e.g., from
interaction with the conduction electron bath.
The master equation reads \cite{Sipe}
\begin{equation}
\label{density matrix}
\dot{\rho} = -\frac{i}{\hbar} [H^{\textrm{eff}},\rho]- \frac{\gamma}{2}\sum_{i}[\sigma_i,[\sigma_i,\rho]],
\end{equation}
where $\gamma$ is the dephasing rate. %Noting that
%the state evolution without any noise or dephasing effects can be described by the master equation (\ref{density matrix}),
%when $\gamma=0$.
We introduce the Bloch vector with components $u=\rho_{1-1}+\rho_{-11}$, $v= -i(\rho_{1-1}-\rho_{-11})$, and $w=\rho_{11}-\rho_{-1-1}$,
and obtain from Eq. (\ref{density matrix})
\beqa
\label{bloch equation}
\left(\begin{array}{ccc}
 \dot{u}
\\
\dot{v}
\\
\dot{w}
\end{array}\right)
=
\left(\begin{array}{ccc}
-4 \gamma & \eta Z & - \eta Y
\\
-\eta Z &-4 \gamma & \eta X
\\
\eta Y  & - \eta X & -4 \gamma
\end{array}\right)
\left(\begin{array}{ccc}
u
\\
v
\\
w
\end{array}\right).
\eeqa

\begin{figure}[]
\scalebox{0.50}[0.50]{\includegraphics{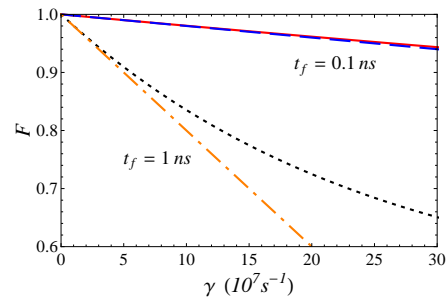}}
\caption{(Color online) Fidelity as a function of $\gamma$ for different times $t_f=0.1$ ns (solid red)
and $t_f=1 $ ns (dotted black). The fidelity estimated from perturbation theory is also compared,
for $t_f=0.1$ ns (dashed blue, undistinguished) and $t_f=1 $ ns (dot-dashed orange).
$B_0=0.15$ T and other parameters are the same as in Fig. \ref{Bmax}.}
\label{fidelity}
\end{figure}

We solve Eq.(\ref{bloch equation}) numerically and calculate fidelity $F = |\la -1  | \Psi (t_f)\ra|$, see Fig. \ref{fidelity}.
%, where $|-1 \ra$ is the target state.}
For $\gamma t_{f}\ll1$ the time-dependent perturbation theory \cite{3d} yields the bound
$
F  \gtrsim 1 - 2 \gamma t_f
$.
%This motivates the Clearly, the fidelity decreases with increasing the time $t_f$ and $\gamma$.  The comparison in Fig. \ref{fidelity} demonstrates that the shorter time is, the more robust
%the process becomes.
Since the induced flip occurs very fast, it can overcome the main danger for
the low-temperature spin manipulation in QDs coming from the hyperfine coupling to
the nuclear spins, where the decoherence times exceed  $100$ ns \cite{Nowack}.
%, which makes the decoherence effect weaker.

%%%%%%%%%%%%%%%%%%%%%%%%%%%%%%%%%%%%%%%%%%%%%%%%%%%%%%%%%%%%%%%%%%%%%%%%%%%%%%%%%%%%%%%%%%%%%%%%%%%%%%%%
%%%%%%%%%%%%%%%%%%%%%%%%%%%%%%%%%%%%%%%%%%%%%%%%%%%%%%%%%%%%%%%%%%%%%%%%%%%%%%%%%%%%%%%%%%%%%%%%%%%%%%%%

%For smaller manipulation time, for example $t_f=10^{-10}s$, as show in , $F_a$ decays slowly and almost coincide with $F_b$ with up to $\Gamma=15\times10^7 $ s$^{-1}$. In contrast, for longer manipulation time $t_f=10\times10^{-10}s$, shown in Fig. \ref{fidelity}(b), both $F_a$ and $F_b$ decay more rapidly than those with $t_f=10^{-10}s$, while they separate with each other from weaker $\Gamma$. s between $F_a$ and $F_b$ with longer and shorter $t_f$ highlight the advantage of our proposal: Since the time duration that spin flip is performed .

%By means of Lewis-Riesenfeld method, the elements of $H'$: $Z$, $X$ and $Y$ result in the variations of fidelity. In this sense, the value of $\xi$, representing the ratio played by the upper two energy levels (See Appendix), has no impacts on the fidelity but leads to the changes of the electric field that we designed. On the other hand, we can make conclusion that effects from other states on the spin flip between any couplet can be characterized
%by $\xi$, as long as we sum up all components from other states, from the perturbation theory.

Another source of decoherence is the device-dependent noise in the electric
field acting on the spin. This can be important when the relatively weak electric fields are applied. We analyze in
detail the effect of this noise and find that our method is robust to this randomness in \cite{suppl material-2}.

\textit{Conclusions and outlook.-}
We have proposed a fast and robust method to flip electron spin
in a QD with SO coupling and weak perpendicular magnetic field.
The spin-flip process, designed by Lewis-Riesenfeld invariants, is faster than the decoherence
for any known low-temperature dephasing mechanism.
This method can be further complemented by optimal control theory for time- and energy-minimization subjected to
different physical constraints \cite{Boscain}. Implementation of this technique will allow for
high-fidelity spin-manipulation for quantum information processing.
%Our theoretical results seem to be promising enough to speed up the adiabatic passage for the quantum state control in semiconductor QDs \cite{Simon,Wu}.

{\it Note added:} We have corrected errors in the published version. 

%\section*{Acknowledgement}
We acknowledge funding by the Basque Government (Grants No.
IT472-10 and BFI-2010-255), Ministerio de Ciencia e Innovacion (Grant No.
FIS2009-12773-C02-01), UPV/EHU program (UFI 11/55),
and National Natural Science Foundation of China (Grant No. 61176118) and
Shanghai Rising-Star Program (Grant No. 12QH1400800).

\newpage

\appendix
\section{I. Hamiltonian reduction}
%\vspace{1cm}

Here we give the details on the derivation of the effective $2 \times 2$ Hamiltonian [Eq. (6)] in the paper.
For simplicity we consider a 4-level system. The total Hamiltonian is $H=H_0+H_{\rm so}+H_{\rm int}$.
In the basis of two lowest
spin-split orbital states,
\begin{widetext}
\beqa
H_0 =
\left(
\begin{array}{cccc}
E_1+\Delta_z/2 & 0 & 0 &0
\\
0 & E_1-\Delta_z/2  & 0  & 0
\\
0 & 0  & E_2+\Delta_z/2  & 0
\\
0 & 0  & 0  & E_2-\Delta_z/2
\end{array}
\right),
\eeqa
where $\Delta_z$ is the Zeeman splitting, and
\beqa
H_{\rm so}+H_{\rm int} =
\left(
\begin{array}{cccc}
-\displaystyle{\frac{e}{c}} \beta A_x & -\displaystyle{\frac{e}{c}} \alpha ( i A_x + A_y) & (\beta -\tilde{A}_x) \overline{p}_x - \tilde{A}_y \overline{p}_y & \alpha (i\overline{p}_x + \overline{p}_y)
\\
\displaystyle{\frac{e}{c}} \alpha (i A_x - A_y) & \displaystyle{\frac{e}{c}} \beta A_x & \alpha (-i\overline{p}_x + \overline{p}_y) & -(\beta +\tilde{A}_x) \overline{p}_x - \tilde{A}_y \overline{p}_y
\\
(-\beta +\tilde{A}_x) \overline{p}_x + \tilde{A}_y \overline{p}_y  & -\alpha (i\overline{p}_x + \overline{p}_y) & -\displaystyle{\frac{e}{c}} \beta A_x & -\displaystyle{\frac{e}{c}} \alpha ( i A_x + A_y)
\\
\alpha (i\overline{p}_x - \overline{p}_y) & (\beta +\tilde{A}_x) \overline{p}_x + \tilde{A}_y \overline{p}_y & \displaystyle{\frac{e}{c}} \alpha (i A_x - A_y) & \displaystyle{\frac{e}{c}} \beta A_x
\end{array}
\right),~~~~
%\end{split}&
\eeqa
\end{widetext}
%\beqa
%\label{H-elements}
%H_{11} &=& -H_{22}=H_{33}=-H_{44}=-\frac{e}{c} \beta A_x,
%\\
%H_{12} &=& H^*_{21}=H_{34}=H^*_{43}= -\frac{e}{c} \alpha ( i A_x + A_y),
%\\
%H_{13} &=& H^*_{31}=(\beta -\frac{e}{mc} A_x) \overline{p}_x - \frac{e}{mc} A_y \overline{p}_y,
%\\
%H_{14} &=& H^*_{41}=-H^*_{23}=-H_{32}=\alpha (i\overline{p}_x + \overline{p}_y),
%\\
%H_{24} &=& H^*_{42}=-(\beta +\frac{e}{m c} A_x) \overline{p}_x - \frac{e}{m c} A_y \overline{p}_y.
%\eeqa
with %the momentum transition from
%$\Psi_2$ to $\Psi_1$ in $x$ and $y$ components,
$\overline{p}_{i} = \la \psi_1 |
%\hat
{p}_{i} | \psi_2 \ra$ and $\tilde{A}_{i}\equiv eA_{i}/mc$. %, respectively.

Our aim is to flip the spin between two levels taking into account the effects from orbital motion and the influence
of the other levels. As the energy gap of the orbital states is much larger than the Zeeman splitting $\Delta_z$, the L\"{o}wdin partition \cite{lowdin-partition,Winkler} will enable us to obtain an effective $2 \times 2$ matrix Hamiltonian. We split
the total Hamiltonian $H$ into four blocks $\hat{Q}$, $\hat{B}$, $\hat{C}$, $\hat{C}^\dagger$, each of which is a $2\times2$ matrix,
\begin{eqnarray}
\label{H BLOCK}
H = \left(\begin{array}{cc} \hat{Q} & \hat{C} \\ \hat{C}^\dagger & \hat{B} \end{array}\right).
\end{eqnarray}
The time-independent Schr\"{o}dinger equation can be
formally written as
%solved through ($\bm{F}$) and ($ \bm{G}$) subspaces, as follows:
%
%\begin{eqnarray}
%\label{H BLOCK}
%\left[\begin{array}{cc} \hat{Q} & \hat{C} \\ \hat{C}^\dagger & \hat{B} \end{array}\right] \left[\begin{array}{c} \bm{F} \\ \bm{G} \end{array}\right]
%= E \left[\begin{array}{c} \bm{F} \\ \bm{G} \end{array}\right].
%\end{eqnarray}
%
%Multiplying the blocks,
%
\begin{eqnarray}
\label{line up}
\hat{Q} \bm{F} + \hat{C} \bm{G} &=& E \bm{F},
%\end{eqnarray}
\\
%\begin{eqnarray}
\label{line down}
\hat{C}^\dag \bm{F} + \hat{B} \bm{G} &=& E \bm{G},
\end{eqnarray}
Substituting the formal solution of
Eq. (\ref{line down}), $\bm{G}=(E-\hat{B})^{-1}C^\dagger \bm{F}$,
into Eq. (\ref{line up}), we get a closed equation for $\bm{F}$,
\begin{eqnarray}
\label{effective F}
\hat{Q} \bm{F} + \hat{C}(E-\hat{B})^{-1}C^\dagger \bm{F} = E \bm{F}.
\end{eqnarray}
Therefore, the effective Hamiltonian including the effects from $\hat{B}$, $\hat{C}$ and $\hat{C}^\dagger$
in this subspace is given by $H^{\textrm{eff}} \rightarrow \hat{Q}+\hat{C}(E-\hat{B})^{-1}C^\dagger.$

Assuming that {the electric field is not strong enough to excite other orbital states,}
namely,
$e\max\left(\left|A_{x}\overline{p}_x\right|,\left|A_{y}\overline{p}_y\right|\right)/m c \ll E_2-E_1$,
and keeping the first order term of the spin-orbit coupling constant,
we obtain the effective Hamiltonian
\begin{eqnarray}
\label{Heff}
H^{\textrm{eff}} = \left(\begin{array}{cc} E_1+ h^0_{11}+h_{11}
&  h^0_{12} +h_{12}
\\  h^0_{21} + h_{21}
&E_1 + h^0_{22}+h_{22}
\end{array}\right),
\end{eqnarray}
where
\beqa
\label{Heff-elements}
 h^0_{11} &=& -h^0_{22}=\Delta_z - \frac{e}{c} \beta A_x,
\\ h^0_{12} &=& h^{0*}_{21} = -\frac{e}{c} \alpha (i A_x + A_y),
\\ h_{11} &=& -h_{22}= -2 \frac{e \beta}{mc} \frac{\overline{p}_x \left(A_x \overline{p}_x + A_y \overline{p}_y\right)}{ E_2-E_1},
\\ h_{12} &=&  h_{21}^* = -2 \frac{e \alpha}{mc} \frac{\left( A_x \overline{p}_x + A_y \overline{p}_y\right) \left( i \overline{p}_x + \overline{p}_y\right)}{E_2-E_1}.
\eeqa
By a simple shift we may ignore the common term $E_1$ on the diagonal, and finally express the Hamiltonian as
%\begin{eqnarray}
%\label{Heffsimplified}
%H^{'}=\left(\begin{array}{cc}  E_1 + \Delta - \beta e A_x (1-\xi) & - i \alpha e A_x (1-\xi) - \alpha e A_y \\
%i \alpha e A_x (1-\xi) - \alpha e A_y  &  E_1 - \Delta + \beta e A_x (1-\xi) \end{array}\right).
%\end{eqnarray}
%Finally, the spin-dependent part is rewritten in the following form:
\beqa
\label{Heff2}
H^{\textrm{eff}} = \frac{g \mu_B}{2}
\left(
\begin{array}{cc}
Z & X + i Y
\\
X - i Y & -Z
\end{array}
\right),
\eeqa
where  $X = B_2(1+\xi_y) $, $Y = (\alpha/\beta)(1+\xi_x) B_1 $, $Z = B_0 + (1+\xi_x) B_1 $
with $B_1=-2e\beta A_x/(c g \mu_B)$, $B_2=-2e\beta A_y/(c g \mu_B)$.
According to perturbation theory, the effects from other states can be characterized
by $\xi_x$ and $\xi_y$, $\xi_i=2 \sum_{n>1}\left|\la \psi_1|
%\hat
{p}_i|\psi_n \ra\right|^2/[m(E_n-E_1)]$, summing up all contributions.
%, where $n>1$
%represents other excited states.
%Moreover, $\xi_x$ and $\xi_y$ only modulate the amplitude
%of the designed electric field with no impact on transition fidelity. ***Could you explain this further? (at least to me)***

\section{II. Apparatus Noise Effect}

Besides the decoherence resulting from the interaction between the system and the environment introduced in the main text,
we consider apparatus noise, i.e. the Hamiltonian $H^{\textrm{eff}}$ [Eq. (\ref{Heff2})] perturbed by a
stochastic part $H^{\textrm{noise}}$ describing noise from the electric field source. Therefore,
the Stochastic Schr\"{o}dinger equation is
\beqa
\label{stochastic Schrodinger equation}
i \hbar \frac{d}{dt} \Psi(t) = (H^{\textrm{eff}} + H^{\textrm{noise}}) \Psi(t).
\eeqa
Here $H^{\textrm{noise}} = \lambda H' \zeta(t)$, and $\lambda = \lambda_0 \sqrt{t_f}$, where $\zeta(t) =d W/dt$ is heuristically the time-derivative of the Brownian motion
$W$ \cite{Zoller}, and $\lambda_0$ is the strength of the noise.
We have $\la \zeta(t) \ra = 0 $ and $ \la \zeta(t) \zeta(t') \ra = \delta(t-t')$ because the noise should have zero mean value and should be uncorrelated at different times. $H'$ is the part of $H^\textrm{eff}$ contributing from the time-dependent electric field, assumed here to be parallel to the $x$-axis for the definiteness. The resulting
short-time evolution of the wave function can be statistically presented
using Ito approach as \cite{amplitude noise}:

\begin{equation}
\label{stochastic Schrodinger equation 1}
| \Psi (t+dt) \ra =e^{-i\left(H^{\textrm{eff}}dt + \lambda H'dW \right)/\hbar}|\Psi (t) \ra,
\end{equation}
where $dt$ is the infinitesimal time step and $dW$ is the corresponding noise increment in the Ito calculus \cite{Zoller}.
The properties of such noise provide $\la dW \ra =0$, $\la d W^2 \ra =dt$. By expanding (\ref{stochastic Schrodinger equation 1}) in Taylor series,
and keeping the terms with first order in $dt$ and $dW$, we obtain the increment
\begin{equation}
\label{SSE}
|d \Psi \ra = - \frac{i} {\hbar} H^{\textrm{eff}} dt |\Psi \ra - \frac{\lambda^2}{2 \hbar^2} H'^2 dt |\Psi \ra - \frac{i \lambda}{\hbar} H' d W |\Psi \ra,
\end{equation}
from which time dependence of density matrix is given by \cite{amplitude noise}
\beqa
\label{density matrix}
\dot{\rho} = -\frac{i}{\hbar} [H^{\textrm{eff}},\rho] - \frac{\lambda^2}{2 \hbar^2} [H',[H',\rho]],
\eeqa
where the second term is responsible for the decrease in the process fidelity due to the apparatus noise.
The master equation (\ref{density matrix}) differentiates
the amplitude-noise effect from the relaxation and decoherence, described by the conventional Lindblad master equation.
%
%%%%%%%%%%%%%%%%%%%%%%%%%%%%%%%%%%%%%%%%%%%%%%%%%%%%%%%%%%%%%%%%%%%%%%%%%%%%%%%%%%%%%%%%%%%%%%%%%%%%%%%%%%%%%%%%%
\begin{figure}[]
\scalebox{0.60}[0.60]{\includegraphics{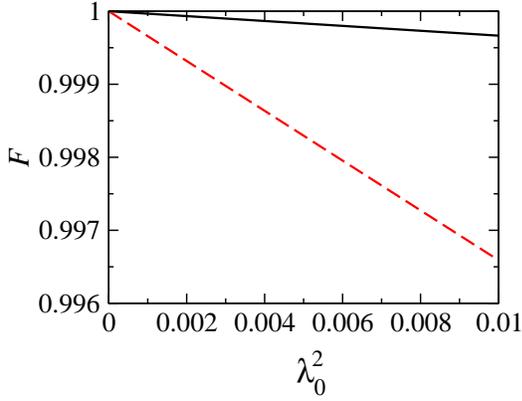}}
\caption{(Color online) Dependence of fidelity on $\lambda^2_0$, given by $t_f = 1$ ns
(solid black) and $t_f = 0.1$ ns (dashed red).
Other parameters are the same as in Fig. 2 (b) in the main text: $\hbar \alpha=2 \times 10^{-6}$ meV$\cdot$cm,
$\hbar \beta = 10^{-6}$ meV$\cdot$cm, $g=-0.44$
for GaAs, $B_0=0.15$ T, $\xi_x =\xi_y = 0$.}
\label{noise}
\end{figure}
%%%%%%%%%%%%%%%%%%%%%%%%%%%%%%%%%%%%%%%%%%%%%%%%%%%%%%%%%%%%%%%%%%%%%%%%%%%%%%%%%%%%%%%%%%%%%%%%%%%%%%%%%%%%%%%%%%

Following the same procedure to derive the Bloch equation as in the main text after Eq. (22),
we obtain the corresponding equation with the source noise. By introducing three-component state vector
$(u,v,w)$ we find the time evolution:
\begin{widetext}
\beqa
\label{Bloch equation}
\left(\begin{array}{ccc}
 \dot{u}
\\
\dot{v}
\\
\dot{w}
\end{array}\right)
=
\left(
\begin{array}{ccc}
-\frac{1}{2} \lambda^2 \eta^2 (Y^2+Z'^2) & \eta Z & -\eta Y
\\
-\eta Z & -\frac{1}{2} \lambda^2 \eta^2 (X^2+Z'^2) & \eta X
\\
\eta Y & -\eta X & -\frac{1}{2} \lambda^2 \eta^2 (X^2+Y^2)
\end{array}
\right)
\left(\begin{array}{ccc}
{u}
\\
{v}
\\
{w}
\end{array}\right),
\eeqa
\end{widetext}
where in addition to the quantities introduced in the main text, $Z'=Z-B_0 = (1+ \xi_x) B_1$.
In order to analyze the effects from randomness induced by the electric field,
we calculate numerically with Eq. (\ref{Bloch equation}) the fidelity $F = |\la -1| \Psi (t_f)|$,
see Fig. \ref{noise}. The high fidelity shows that our method is robust to the source noise.

\end{document}